\documentclass[seceq]{ptptex}
\usepackage{wrapft}

\usepackage{graphicx}

\title{Long-Range Spectral Statistics of Classically Integrable Systems}
\subtitle{Investigation along the Line of the Berry-Robnik Approach}

\author{
Hironori \textsc{Makino}
$^{1,}$\footnote{E-mail: makino@tokai.ac.jp} 
and Shuichi \textsc{Tasaki}$^{2}$
}

\inst{
$^1$Department of Human and Information Science, 
Tokai University, Hiratsuka 259-1292, Japan
\\
$^2$Department of Applied Physics, Waseda University, 
Tokyo 169-8555, Japan
}

\recdate{%      %Editorial Office will fill in this.
April 25, 2005}

\abst{
Extending the argument of Ref.\citen{[4]} to the long-range spectral statistics of classically integrable quantum systems, we examine the level number variance, spectral rigidity and two-level cluster function.  These observables are obtained by applying the approach of Berry and Robnik\cite{[0]} and the mathematical framework of Pandey \cite{[2]} to systems with infinitely many components, and they are parameterized by a single function $\bar{c}$, where $\bar{c}=0$ corresponds to 
Poisson statistics, and $\bar{c}\not=0$ indicates deviations from Poisson statistics.  This implies that even when the spectral components are statistically independent, non-Poissonian spectral statistics are possible.
}

\begin{document}
\maketitle

\section{Introduction}

\label{sect1}
For bounded quantum systems in the semiclassical limit, the statistical properties of eigenenergy levels have been studied intensively.  Universal behavior has been found in the statistics of {\it unfolded} energy levels \cite{[3]}, which are the sequences of numbers uniquely determined by the mean level density in a small energy interval obtained using the Thomas-Fermi rule.  It is widely known that for quantum systems whose classical counterparts are strongly chaotic, the energy level statistics are well characterized by random matrix theory (RMT), in accordance with the conjecture of Bohigas et al.\cite{[3.5]}, which gives a Wigner-type level-spacing distribution.  For systems whose classical counterparts are integrable, the energy level statistics are well characterized by Poisson statistics in accordance with the conjecture of Berry and Tabor\cite{[6]}, which gives a Poisson-type level-spacing distribution.

Universal behaviors are also exhibited by other quantities, e.g., the level number variance $\Sigma^2(L)$, which is the average variance of the number of levels in an interval containing an average of $L$ levels \cite{[7]}.  For an unfolded level sequence, this interval is equivalent to an interval of length $L$, and $\Sigma^2(L)$ is thus defined as
\begin{equation}
\Sigma^2(L)=\langle\left({\mathcal{N}}(E+L)-{\mathcal{N}}(E)-L\right)^2\rangle, 
\label{1.0}
\end{equation}
where ${\mathcal{N}}(E)$ is the number of eigenenergy levels below $E$, and the brackets $\langle\cdots\rangle$ represent the average over the value of $E$.  The number variance depends primarily on the long-range correlation between the energy eigenvalues, in contrast to the level spacing distribution, which depends primarily on the short-range correlation.  In RMT, for which $\beta=1,2,$ and $4$ correspond, respectively, to GOE, GUE, and GSE level statistics \cite{[7]}, the number variance increases logarithmically with $L$ $\left[\mbox{explicitly, }\Sigma^2(L)\sim (2/\beta\pi^2)\log(2\pi L) \right]$, while in Poisson statistics, the number variance is equal to the number itself $\left[\mbox{i.e., }\Sigma^2(L)=L\right]$.  The latter behavior follows from the fact that $\Sigma^2(L)$ can be expressed in terms of the two-level cluster function \cite{[28]} $Y_2(r)$ as
\begin{equation}
\Sigma^2(L) =L - 2\int_0^L (L-r)Y_2(r)dr,
\label{1.1}
\end{equation}
where $Y_2(r)$ is ordinarily defined in terms of the two-level correlation function \cite{[8]}.  In the absence of spectral correlations, we have $Y_2(r)=0$ for all $r$, which implies $\Sigma^2(L)=L$.

The level statistics for integrable quantum systems have been theoretically studied in a number of works\cite{[6],[9],[10],[11],[12],[13],[14],[4],[15],[Eskin],[Sarnack]}, and have been the subject of many numerical investigations.  In Refs.\citen{[15]} and \citen{[Eskin]}, eigenvalue problems for integrable quantum systems are reformulated as lattice point problems.  There, it is proved, under explicit diophantine conditions, that the two-level correlation function exhibits Poisson statistics.

The appearance of Poisson statistics is now widely acknowledged to be a universal phenomenon in generic integrable quantum systems.  However, the mechanism responsible for this statistics is still not well understood, and deviations from Poisson statistics have also been found.\cite{[4],[6],[12],[26],[27],[14],[15],[Eskin]}

As suggested, e.g., by Hannay (see the discussion given in Ref.\citen{[0]}), one possible explanation for the appearance of Poisson statistics is the following: For an integrable system of $f$ degrees of freedom, almost every orbit is generically confined in its inherent torus, and the entire region in the phase space is densely covered by invariant tori, as suggested by the Liouville-Arnold theorem.\cite{[16]}\  In other words, the phase space of an integrable system consists of infinitely many tori that have infinitesimal volumes with respect to the Liouville measure.  Thus, the energy level sequence for the whole system consists of the superposition of the subsequences that are contributed from those regions.  Therefore, if the mean level spacing of each independent subset is large, one would expect Poisson statistics as a result of the law of small numbers.\cite{[17]}   This explanation given by Hannay is supported by the theory proposed by Berry and Robnik.\cite{[0]}

The Berry-Robnik theory relates the formation of the eigenenergy levels to the phase-space geometry by assuming that the sequence of the entire energy spectrum is given by the superposition of statistically independent subspectra, which are contributed respectively from eigenstates localized in the invariant regions in phase space.  The existence of such independent subspectra is a consequence of the condensation of energy eigenfunctions on disjoint regions in the classical phase space and the resulting lack of overlap among these eigenfunctions.  Therefore, independent subspectra can be expected only in the semiclassical limit, in which the Planck constant tends to zero $(\hbar\to 0)$.  This mechanism was first proposed by Li and Robnik \cite{[18]} on the basis of an implicit state described by Berry \cite{[19],[20]}, and it is sometimes referred to as {\it{the principle of uniform semiclassical condensation of eigenstates}} \cite{[21]}.  This principle states that the Wigner function of a semiclassical eigenstate is connected on a region in phase space explored by a typical trajectory of the classical dynamical system.  In integrable systems, the phase space is folded into invariant tori, and the Wigner functions of the eigenstates tend to delta functions on these tori in the semiclassical limit \cite{[20]}.  On the other hand, in a strongly chaotic system, almost all trajectories cover the energy shell uniformly, and hence the Wigner functions of eigenstates are expected to become delta functions on the energy shell, as suggested by the quantum ergodicity theorem \cite{[22],[23]}.  Because of the suppression of tunneling, each quantum eigenstate is folded into independent subsets in the semiclassical limit, and is expected to form independent spectral components.  Indeed, the formation of such independent components has been confirmed numerically in a deep semiclassical regime\cite{[24],[25]}.  

Employing the above point of view, Seligman and Verbaarschot investigated a semiclassical formula for the number variance $\Sigma^2(L)$ in the case that the energy spectrum consists of the superposition  of subspectra contributed by the phase space components.\cite{[1]}\  They introduced the mathematical framework formulated by Pandey\cite{[2]}\ and observed that the long-range spectral statistics of systems consisting of $N$ independent components described by GOE statistics of equal strength approach Poisson statistics in the limit $N\to+\infty$.  

The level statistics of systems with infinitely many spectral components have also been studied in Refs. \citen{[7],[2],[0],[4]} and \citen{[1]}.  For example, in Ref. \citen{[4]}, the level spacing distribution in the limit $N\to+\infty$ is investigated, and it is classified into three cases: (i) Poisson distribution, (ii) Poisson distribution for large level spacings, but possibly not for small spacings, and (iii) sub-Poisson distribution.  It is observed that cases (ii) and (iii) are possible when the spectral components exhibit strong accumulation and have a singular statistical property.

These works seem to imply that the existence of infinitely many independent components is an essential ingredient for the appearance of Poisson statistics and that the semiclassical limit is one situation in which an effectively infinite number of components arises.  However, in some classically integrable systems in which one might expect there to be infinitely many components, deviation from Poisson statistics has been observed. \cite{[4],[6],[12],[26],[27],[14],[15],[Eskin]}\  Therefore, it is interesting to explore the consequences of the existence of infinitely many independent components.

Here we briefly review the formula proposed by Pandey.\cite{[2]}\  Consider a system whose classical phase space is decomposed into $N$-disjoint regions which provide independent spectral components.  When the entire level sequence is a product of a statistically independent superposition of $N$ subsequences, the level number variance for the entire sequence, @$\Sigma^2$, can be decomposed into those of the subsequences, $\Sigma^2_i$, as
%
%
%
%%%%%%%%
\begin{equation}
\Sigma^2(L;N)= \sum_{i=1}^N \Sigma^2_i(\rho_i L),
\label{1.2}
\end{equation}
%%%%%%%%
%
where $\rho_i (i=1,2,3,\cdots,N)$ denotes the statistical weights of the individual spectral components.  This quantity is defined by the mean level spacing $\bar{S}_i$ of the unfolded spectral components as $\rho_i=1/\bar{S}_i$ and satisfies $\sum_{i=1}^N\rho_i=1$.   In terms of the cluster function $Y_{2,i}(r)$ of a subsequence,  $\Sigma^2_i(\rho_i L)$ is given by
%
%%%%%%%%
\begin{equation}
\Sigma^2_i(\rho_i L) = \rho_i L 
-2 \int_0^{\rho_i L}(\rho_i L-r)Y_{2,i}(r)dr,
\label{1.3}
\end{equation}
%%%%%%%%
%
%
where $Y_{2,i}(r)$ satisfies the relation $\int_{-\infty}^{+\infty}Y_{2,i}(r)dr=1$.  

In combination with the number variance, the spectral rigidity $\Delta_3(L)$ introduced by Dyson and Mehta has played a major role in the study of long-range spectral statistics. \cite{[28]}\  It is defined as the mean-square deviation of $N(E)$ from the best-fitting straight line;
\begin{equation}
\Delta_3(L) = \frac{1}{L} \left< 
\min_{A,B}\int_{E_s}^{E_s+L} dE \left( N(E)-AE-B \right)^2 \right>.
\label{1.4}
\end{equation}
The angular brackets $\langle\cdots\rangle$ represent the average over the value of $E_s$.  Because the number variance is connected to the spectral rigidity as \cite{[2]}
\begin{equation}
\Delta_3(L) = \frac{2}{L^4}\int_0^L dr(L^3-2 L^2 r +r^3)\Sigma^2(r),
\label{1.5}
\end{equation}
$\Delta_3(L)$ is also additive and the formula (\ref{1.2}) can be rewritten as
\begin{equation}
\Delta_3(L;N) =\sum_{i=1}^N \Delta_{3,i}(\rho_i L),
\label{1.6}
\end{equation}
where the rigidity of the subspectrum is described by the number variance of the subspectrum as
\begin{equation}
\Delta_{3,i}(\rho_i L) = \frac{2}{L^4}\int_0^L dr
(L^3-2 L^2 r +r^3)\Sigma_i^2(\rho_i r).
\label{1.7}
\end{equation}
In the argument of Berry and Robnik,\cite{[0]}\ the statistical weights of the individual components in the formulae ({\ref{1.2}}) and ({\ref{1.6}}) are equivalent to the phase volumes (Liouville measures) of the corresponding invariant regions.  This relation is satisfactory in a given small energy interval if the Thomas-Fermi rule for the individual phase space regions still holds. \cite{[4]}\  Therefore, the formulae ({\ref{1.2}}) and ({\ref{1.6}}) relate the level statistics in the semiclassical limit to the phase-space geometry.

In most general cases, the cluster function might be singular.  In such a case, it is convenient to use its cumulative distribution function $c_i$:
\begin{equation}
c_i(L;\rho_i)=2\int_0^{\rho_i L} Y_{2,i}(r)dr.
\label{1.8}
\end{equation}
In addition to Eqs. ({\ref{1.2}}) and ({\ref{1.3}}), we assume the following two conditions for the statistical weights: 
\begin{itemize}
\item Assumption (i): The statistical weights of individual components uniformly vanish in the limit of infinitely many regions:
\begin{equation}
\max_i \rho_i \to 0\quad\mbox{as}\quad N\to +\infty.
\label{1.9}
\end{equation}
\item Assumption (ii): The weighted mean of the cumulative cluster function,
\begin{equation}
c(r;N)=\sum_{i=1}^N \rho_i c_i(r;\rho_i),
\label{1.10}
\end{equation}
converges in the $N\to +\infty$ limit to $\bar{c}(r)$:
\begin{equation}
\lim_{N\to +\infty}c(r;N) = \bar{c}(r).
\label{1.11}
\end{equation}
The limit is uniform on each closed interval in the range $0\le r \le L$.
\end{itemize}

Under Assumptions (i) and (ii), formula ({\ref{1.2}}) leads to the 
following formulae in the limit $N\to +\infty$:
\begin{eqnarray}
\Sigma^2_{\bar{c}}(L)&=& L-\int_0^L \bar{c}(r)dr,\\
\label{1.12}
\Delta_{3,\bar{c}}(L)&=&\frac{L}{15}+\frac{2}{L^4}\int_0^L (L^3-2L^2x+x^3)\int_0^x dr\bar{c}(r)
\label{1.13}.
\end{eqnarray}
When the mean level spacings of individual components are sufficiently sparse $(\mbox{i.e.,} \bar{S}_i\to+\infty),$ we expect that $\bar{c}(r)=0$, and the limiting observables (\ref{1.12}) and (\ref{1.13}) of the whole energy sequence reduce to those of Poisson statistics, $\Sigma^2_{\bar{c}=0}=L$ and $\Delta_{3,\bar{c}=0}=L/15$.  In general, we expect $\bar{c}(r)\not=0$, which corresponds to a certain clustering (accumulation) of the levels of individual components. 

In the next section, the above statement is proved, and $\Sigma^2(L;N)$ and $\Delta_3(L;N)$ in the limit $N\to+\infty$ are also investigated.  In section \ref{sect3}, the cluster function $Y_2$ of the whole energy levels is also analyzed in the $N\to+\infty$ limit.  We give one possible example of the deviation from Poisson statistics in section \ref{sect4}, and in the concluding section, we discuss some relationships between our results and those of related works.

\section{Number variance and spectral rigidity}
\label{sect2}
In this section, starting from Eqs. ({\ref{1.2}}) and ({\ref{1.3}}) and Assumptions (i) and (ii) introduced in the previous section, we show that, in the limit of infinitely many components $(N\to +\infty)$, $\Sigma^2(L;N)$ and $\Delta_3(L;N)$ converge to the formulae ({\ref{1.12}}) and ({\ref{1.13}}), respectively,  with the cumulative cluster function $\bar{c}$.  The convergence is shown as follows.

First we rewrite $\Sigma^2(L;N)$ in terms of the cluster function $Y_{2,i}$ of independent components:
\begin{equation}
\Sigma^2(L;N)= L - 2 \sum_{i=1}^N\int_0^{\rho_i L} (\rho_i L-r)Y_{2,i}(r)dr.
\label{2.1}
\end{equation}
Now, note that using Eq. ({\ref{1.8}}) and integration by parts, we have
\begin{equation}
2\int_0^{\rho_i L}(\rho_i L - r) Y_{2,i}(r)dr=\int_0^L \rho_i c_i(r;\rho_i)dr.
\label{2.2}
\end{equation}
Since the convergence $\sum_{i=1}^N\rho_{i}c_i(r;\rho_i)\to\bar{c}(r)$ as $N\to+\infty$ is uniform on each interval $r\in [0,L]$ by Assumption (ii), $\Sigma^2(L;N)$ has the following limit:
\begin{eqnarray}
\Sigma^2(L;N) &=& L-\int_0^L \sum_{i=1}^N \rho_i c_i(r;\rho_i) dr\label{2.3}\\
&&\longrightarrow L-\int_0^L \bar{c}(r)dr\quad\mbox{as}\ N\to+\infty.\label{2.4}
\end{eqnarray}
Therefore, we obtain the desired results:
\begin{eqnarray}
\lim_{N\to+\infty}\Sigma^2(L;N)&\equiv&\Sigma^2_{\bar{c}}(L) =L-\int_0^L \bar{c}(r)dr,
\label{2.5}
\end{eqnarray}
\begin{eqnarray}
\Delta_{3,\bar{c}}(L) 
&=& \frac{2}{L^4}\int_0^L dr(L^3-2 L^2 r +r^3)\lim_{N\to+\infty}\Sigma^2(r;N)\nonumber\\
&=& \frac{L}{15} +\frac{2}{L^4}\int_0^L (L^3-2L^2r+r^3)\int_0^r dx\bar{c}(x)\label{2.6}.
\end{eqnarray}

We have Poisson statistics, i.e., $\Sigma^2_{\bar{c}}=L$ and $\Delta_{3,\bar{c}}=L/15$, if the energy levels are uncorrelated, and thus $\bar{c}(r)=0$ for 
all $r$.  Such a case is expected when the individual cumulative function $c_i(r;\rho_i)$ is bounded 
by a finite positive function $g(r)$ as
\begin{equation}
\left| c_i(r;\rho_i)\right|\leq \rho_i^{\eta} g(r),
\label{2.7}
\end{equation}
for all $i$ and $\eta >0$.  Indeed, we have
\begin{eqnarray}
\left| c(r;N)\right|\leq\sum_{i=1}^N \rho_i\left| c_i(r;\rho_i)\right|
\leq g(r)\sum_i^N \rho_i\rho_i^{\eta}\leq g(r)(\max_i\rho_i)^{\eta}
\longrightarrow 0\equiv\bar{c}(r),\quad\mbox{as}\quad N\to\infty.
\label{2.8}
\end{eqnarray}
More specifically, for example, we have $\bar{c}=0$ if the absolute values of individual cluster functions are uniformly bounded by a positive constant $D$: $|Y_{2,i}(r)|\le D$ ($1\le i \le N$). Indeed, the following holds:
\begin{eqnarray}
| c(L;N) |&=& \left | 2\sum_{i=1}^N\rho_i 
\int_{0}^{\rho_i L} Y_{2,i}(r) dr \right|\nonumber\\
&& \leq 2DL\sum_{i=1}^N\rho_i^2\leq 2DL\max_i\rho_i\sum_{i=1}^N
\rho_i\longrightarrow 0\equiv {\bar c}(L),
\quad\mbox{as}\quad N\to+\infty.
\label{2.9}
\end{eqnarray}
This includes the case in which the overall energy spectrum consists of the superposition of $N$ chaotic components characterized by GUE level statistics, $Y_{2,i}(r)=\left( \frac{\sin{(\pi r)}}{\pi r} \right)^2$.  Thus, we have
\begin{eqnarray}
c(L;N) &=&2\sum_{i=1}^N \int_0^L (L-r)dr \left( \frac{\sin{(\pi\rho_i r)}}{\pi r} \right)^2\nonumber\\
&&\leq 2L^2\sum_{i=1}^N\rho_i^2 \nonumber\\
&&\leq  L^2 \max_i \rho_i\sum_{i=1}^N\rho_i\rightarrow 0 \quad\mbox{as}\quad N\to+\infty,
\label{2.10}
\end{eqnarray}
where we have used the relation $\sin{\sigma}\leq\sigma$ for 
$\sigma\geq0$ and Assumption (i).

In general, we expect that $\bar{c}(L)\not=0$, which corresponds to a strong clustering of the levels of individual components, leading to a singular cluster function $Y_{2,i}(L)$ and a nonsmooth cumulative function $c_i(L;\rho_i)$.  Because of the singularity of $Y_{2,i}$ in Eq. (\ref{1.8}), $c_i$ might converge to a nonzero constant even in the limit $N\to+\infty$.  For example, if all $c_i(r)$ simultaneously satisfy $c_i(r)>C>0$ or $c_i(r)<-C<0$ at some two-level distance $r$ with a positive constant $C$, there is deviation from Poisson statistics, and we have $|\bar{c}(r)|=\lim_{N\to+\infty}\sum_{i=1}^N \rho_i |c_i(r)|>C>0$.  Such a strong clustering would be expected when there is a symmetry in the system.  One example of a system that might possess such a symmetry is a rectangular billiard system.  In section \ref{sect4}, we show that such systems exhibit non-Poisson level statistics with $\bar{c}(L)\not=0$.
\section{Two-level cluster function}
\label{sect3}
In this section, we discuss the two-level cluster function of the entire level sequence,
\begin{equation}
Y_2(L;N) = -\frac{d^2}{dL^2}\Sigma^2(L;N),
\label{3.1}
\end{equation}
in the limit of infinitely many independent components, $N\to+\infty$.  

Substituting Eqs. (\ref{1.2}) and (\ref{1.3}) into the above equation, Pandey's formula (\ref{1.2}) can be rewritten for the cluster function as
\begin{equation}
Y_2(L;N)= -\frac{d^2}{dL^2}\sum_{i=1}^N\Sigma^2_i(\rho_i L)= 2\sum_{i=1}^N \rho_i^2 Y_{2,i}(\rho_i L).
\label{3.2}
\end{equation}
The second equation follows from the commutative relation between the sum and the differentiations.  However, in the limit $N\to+\infty$, this relation does not generally hold, due to singularities of the sum.  Therefore, in the present work, we investigate the limit of the two-level cluster function only for the case of the weak convergence to $Y_{2,\bar{c}}(L)$, i.e., the case in which
\begin{equation}
Y_{2,\bar{c}}(L)=-\frac{d^2}{dL^2}\Sigma^2_{\bar{c}}(L).
\label{3.3}
\end{equation}
According to Helly's theorem, \cite{[17],[29]}\ the weak convergence limit is defined by
\begin{equation}
\lim_{N\to+\infty}\int_0^L Y_2(r;N)dr = \int_0^L Y_{2,\bar{c}}(r)dr.
\label{3.4}
\end{equation}
This equation is equivalent to
$\lim_{N\to+\infty}\frac{d}{dL}\Sigma^2(L;N)
=\frac{d}{dL}\Sigma^2_{\bar{c}}(L)$, 
and it can be proved by using Eqs. ({\ref{2.1}}) and ({\ref{2.2}}) and Assumption (ii) as follows:
\begin{eqnarray}
\frac{d}{dL}\Sigma^2(L;N)
&=& \frac{d}{dL}\sum_{i=1}^N\Sigma_i^2(\rho_i L)\nonumber\\
&=& 1-\sum_{i=1}^N \rho_i c_i(L;\rho_i)\label{3.5}\\
&&\longrightarrow 1-\bar{c}(L)=\frac{d}{dL}\Sigma^2_{\bar{c}}(L)
\quad\mbox{as}\quad N\to +\infty.
\label{3.6}
\end{eqnarray}
When the limiting function ${\bar c}(L)$ is differentiable, 
the asymptotic cluster function is given as follows:
\begin{equation}
Y_{2,\bar{c}}(L) = \frac{d}{dL}\bar{c}(L)
\label{3.7}
\end{equation}
\section{Rectangular billiard systems}
\label{sect4}
As an example of deviation from Poisson statistics, we examine a rectangular billiard system whose energy levels are given by
\begin{equation}
\epsilon_{n,m} =  n^2 + \alpha\  m^2,
\label{eq:e}
\end{equation}
where $n$ and $m$ are positive integers, and $\alpha$ is the aspect ratio of the rectangle.  The unfolding transformation of the eigenenergy levels $\{\epsilon_{n,m}\}\to\{\bar{\epsilon}_{n,m}\}$ is carried out by using the leading Weyl term of the cumulative mean number ${\mathcal{N}}(\epsilon)$ of the energy levels,
\begin{equation}
\bar{\epsilon}_{n,m}\equiv {\mathcal{N}}(\epsilon_{n,m}) 
=\frac{\pi}{4\sqrt{\mathstrut\alpha}}\epsilon_{n,m}.
\end{equation}

It is well known that the two-point correlation function exhibits Poisson statistics when $\alpha$ is an irrational number and satisfies the diophantine condition.  This fact was first pointed out by Sarnack\cite{[Sarnack]} and was later proved by Eskin et al. in Ref.\ \citen{[Eskin]}.  While the two-point correlation function is unbounded and does not exhibit Poisson statistics when $\alpha$ is a rational number, expressed as $\alpha=p/q$, where $p$ and $q$ are coprime positive integers.\cite{[Eskin],[15]}\  Thus it is interesting to study the spectral statistics for rational values of $\alpha$.

\begin{figure}
\centerline{\includegraphics[width= 6.8cm,height= 7.5cm]{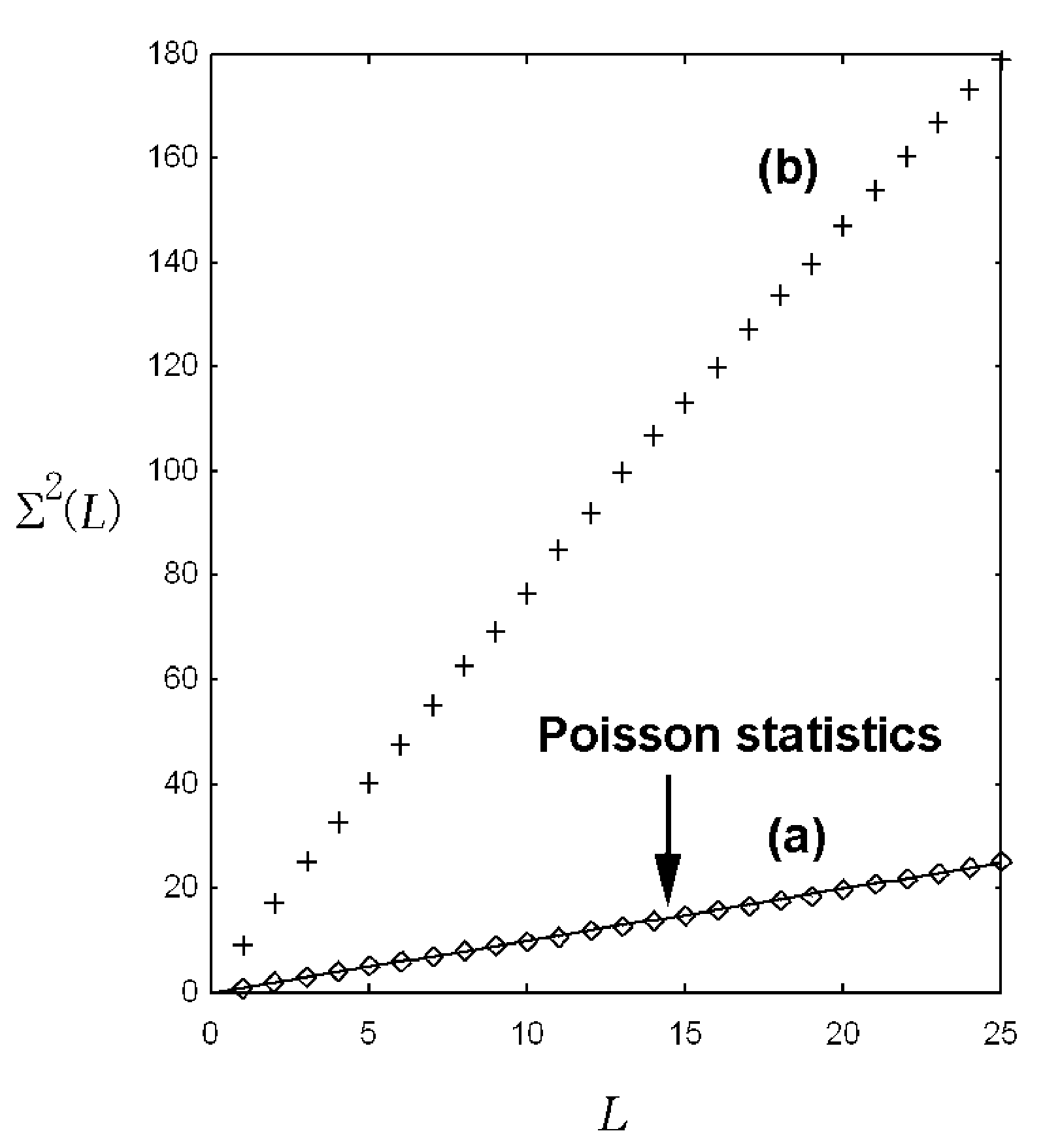}}
\caption{
Numerical results for the level number variance $\Sigma^2(L)$ with (a) 
$\alpha=(1+\sqrt{5})/2$ and (b) $\alpha=1$.  We used 
energy levels $\bar{\epsilon}_{m,i}\in [1000000\times10^6,1000005\times10^6]$.  
The total numbers of levels are (a) 4999087 and (b) 4999224.  The solid 
line represents the number variance in the case of the Poisson statistics, 
$\Sigma^2(L)=L$.}
\label{fig1}
\end{figure}

Fig. \ref{fig1} displays the numerical results of the number variance $\Sigma^2$ for rational and irrational values of $\alpha$.  The numerical computation was carried out using a double precision real number operation.  In the case that $\alpha$ is an irrational number, e.g., the golden ratio, $\alpha=(1+\sqrt{\mathstrut 5})/2$, the number variance exhibits Poisson statistics: $\Sigma^2(L)=L$ [see Fig. \ref{fig1}(a)].  By contrast, in the case that $\alpha$ is a rational number, e.g., $\alpha=1$, a strong clustering of levels appears, due to the corresponding geometrical symmetry of the system, and $\Sigma^2(L)$ clearly deviates from Poisson statistics [see Fig. \ref{fig1}(b)].

\begin{figure}
\centerline{\includegraphics[width=11cm]{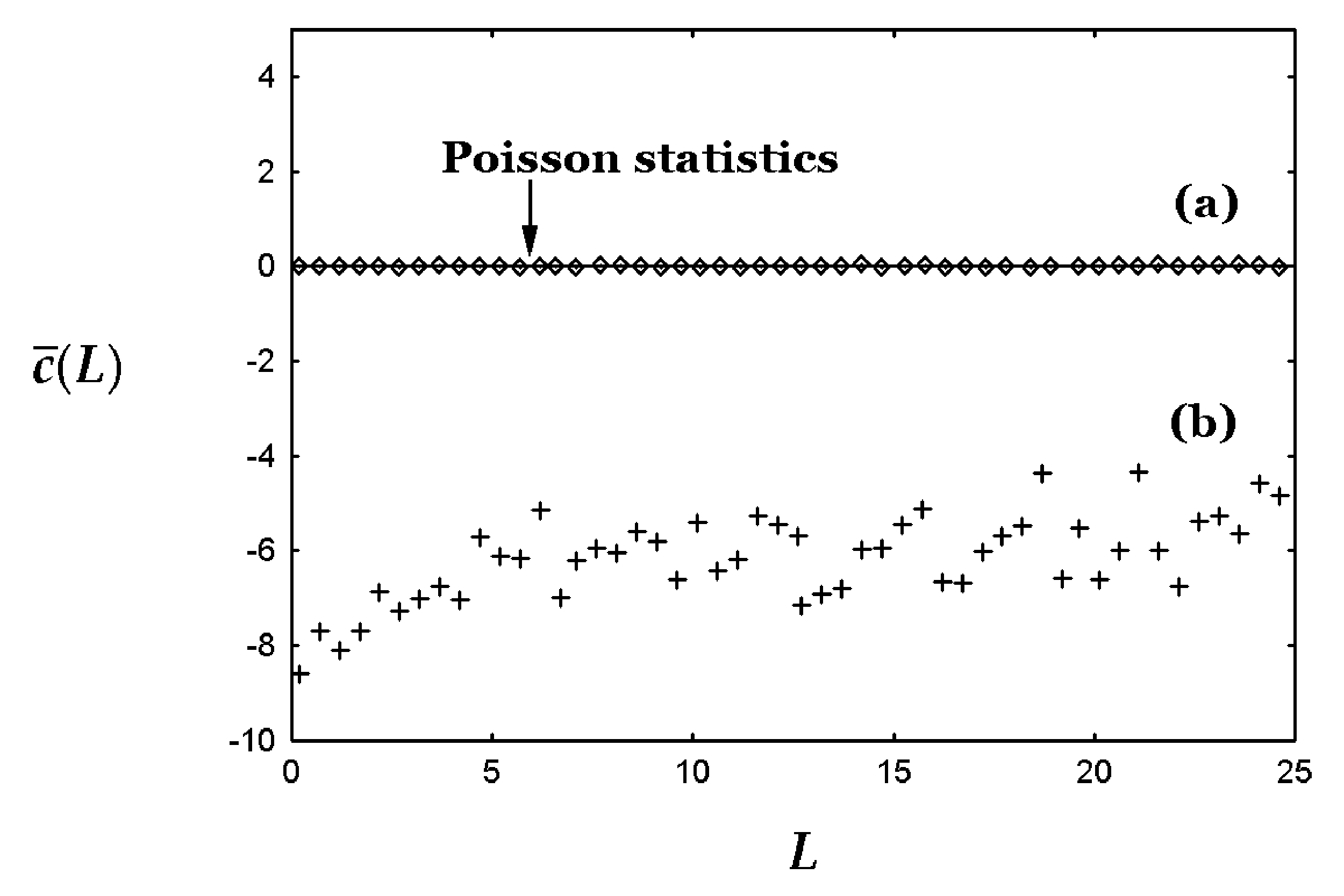}}
\caption{Numerical results for 
$\bar{c}(L)$ with (a) $\alpha=(1+\sqrt{5})/2$ and (b) $\alpha = 1$.  
The solid line represents the number variance in the case of 
Poisson statistics, $\bar{c}(L)=0$.}
\label{fig2}
\end{figure}

Fig. \ref{fig2} displays the numerical results of $\bar{c}(L)$ for the values of $\alpha$ corresponding to Figs. \ref{fig1}(a) and (b), respectively.  For each value of $\alpha$, $\bar{c}(L)$ is calculated from the overall spectrum as
\begin{equation}
\bar{c}(L)=1-\frac{d}{dL}\Sigma^2(L).
\end{equation}
From this, it is clearly observed that $\bar{c}(L)=0$ when $\Sigma^2(L)$ exhibits Poisson statistics, while $\bar{c}(L)\not=0$ when $\Sigma^2(L)$ deviates from Poisson statistics.

\begin{figure}
\centerline{\includegraphics[width=9.7 cm,height=8.3 cm]{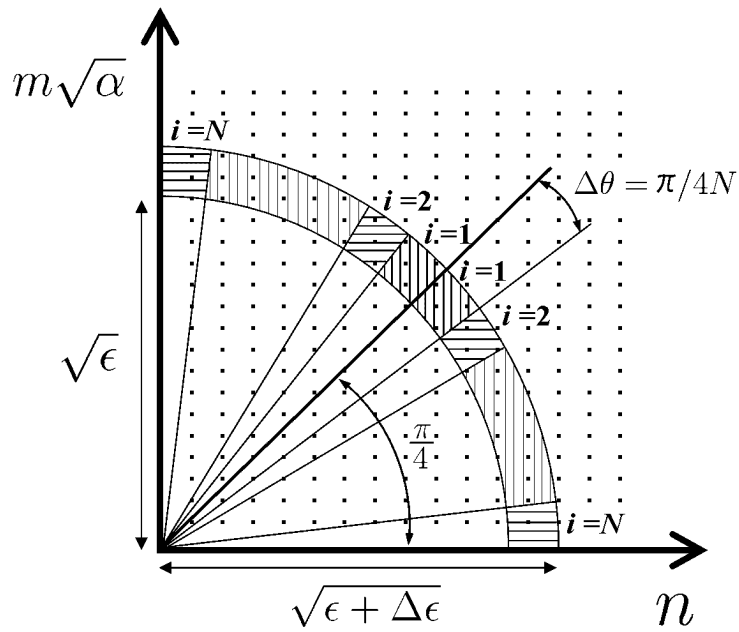}}
\caption{
Schematic picture of identification method for eigenenergy levels in the $(n,m\sqrt{\alpha})$ plane.}
\label{fig3}
\end{figure}

In order to characterize the deviation from Poisson statistics, i.e., $\bar{c}(L)\not=0$ behavior, in terms of the singular statistical properties of the spectral components, we identify each energy level with a component as follows.  Consider a set of eigenstates in the $(n,m\sqrt{\mathstrut\alpha})$ plane, as shown in Fig. \ref{fig3}, where the energy interval $[\bar{\epsilon}, \bar{\epsilon}+\Delta\bar{\epsilon}]$ is represented by the striped region.  This interval is then divided into $2N$ subintervals, each corresponds to an angle of $\Delta\theta=\pi/4N$.  The subintervals are numbered from $1$ to $N$ beginning in the middle of the interval and moving both clockwise and counterclockwise. Here, there are two subintervals corresponding to each number.  Then, the energy level $\bar{\epsilon}_{n,m}\in[\bar{\epsilon},\bar{\epsilon}+\Delta\bar{\epsilon}]$ belonging to each of the subintervals is identified with the $i$-th spectral component as
\begin{equation}
i = \Biggl[ \left|\frac{4N}{\pi}\arctan
\left(\frac{m}{n}\sqrt{\mathstrut\alpha}\right)-N\right|\Biggl]+1,
\label{classify}
\end{equation}
where $[x]$ represents the maximum integer that does not exceed $x$.  With this identification rule, the relative weight of each spectral component, $\rho_i\ (i=1,2,3,\cdots,N)$, is given by
\begin{equation}
\rho_i=\frac{1}{N}\label{ddd},
\end{equation}
where the number of components, $N$, is given in terms of the constant values $\gamma\equiv\Delta\epsilon/\epsilon=\Delta\bar{\epsilon}/\bar{\epsilon}$ and $n_i$ (the number of levels identified with $i$-th component) as
\begin{equation}
N= \Biggl[ \frac{\gamma}{n_i}\bar{\epsilon}  \Biggl].
\end{equation}
The limit of infinitely many components, $N\to+\infty$, corresponds to the high energy limit $\bar{\epsilon}\to+\infty$, which is equivalent to the semiclassical limit.  In this limit, we have $\Delta\theta\to0$, and $\rho_i$ satisfies Assumption (i).

\begin{figure}
\centerline{\includegraphics[width=11 cm]{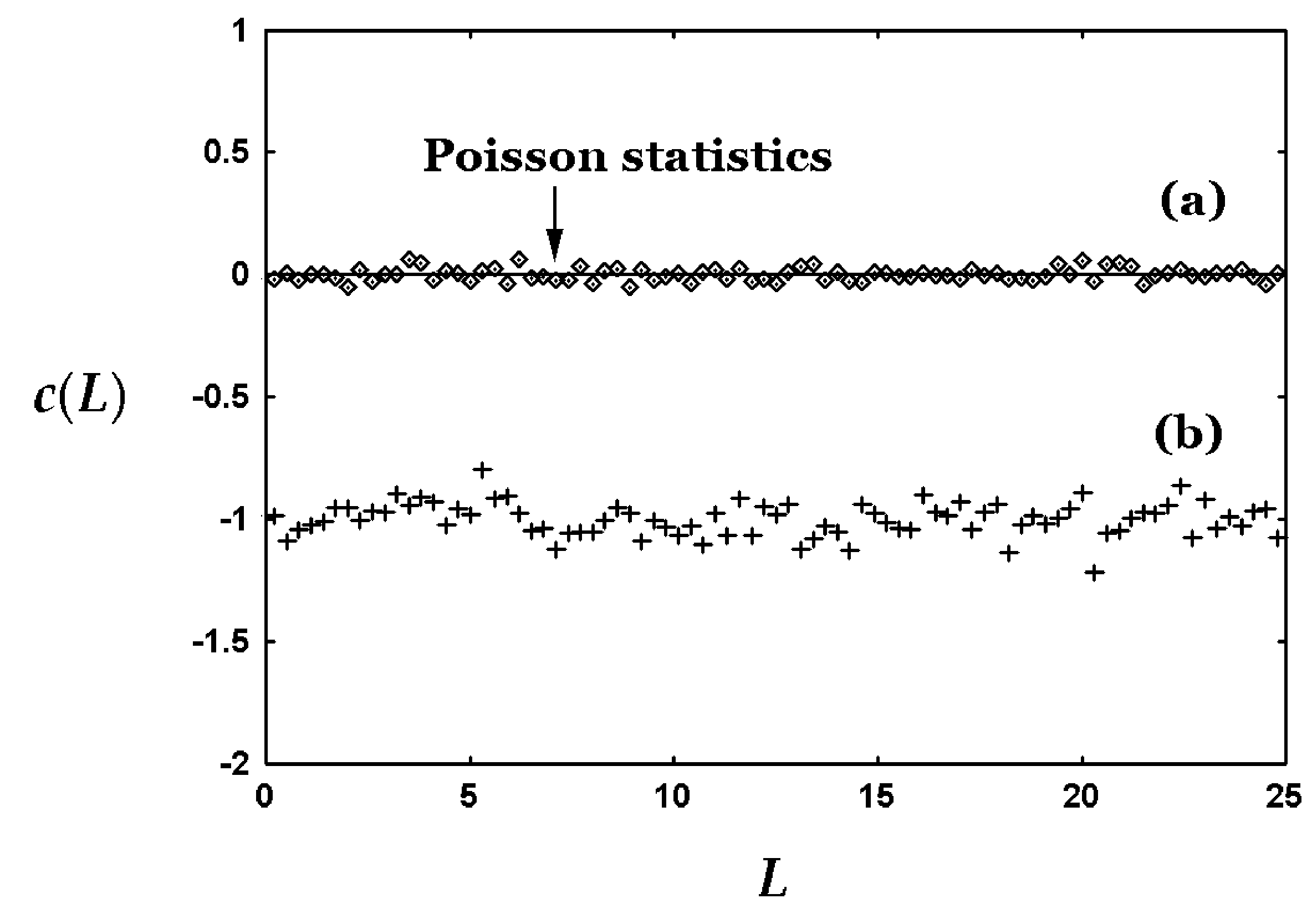}}
\caption{Numerical results for $c(L)$ with 
(a) $\alpha=(1+\sqrt{5})/2$ and (b) $\alpha=1$.  We used energy levels 
$\bar{\epsilon}_{n,m}\in [1000000\times10^6,1000005\times10^6]$.  
The total numbers of levels are 
(a) 4999087 and (b) 4999224, and the number of 
spectral components is $N=1000$.}
\label{fig4}
\end{figure}

Figs. \ref{fig4}(a) and (b) display the numerical results of $c(L)=\sum_{i=1}^N \rho_i c_i(L)$ for the values of $\alpha$ corresponding to Figs. \ref{fig2}(a) and (b), respectively.  In each case, $c_i(L)$ is calculated from the number variance, $\Sigma_i^2$, of the spectral component as follows:
\begin{equation}
c_i(L) = 1-\frac{1}{\rho_i}\frac{d}{dL}\Sigma_i^2(\rho_i L).
\end{equation}
From this, it is clearly seen that $c(L)\not=0$ when $\alpha=1$ and $c(L)=0$ when $\alpha=(1+\sqrt{\mathstrut 5})/2$.  However, as shown in Fig. \ref{fig2}(b), $c(L)$ does not reproduce $\bar{c}(L)$ well, i.e., the deviation from Poisson statistics is not as large as that found for the value of $\bar{c}(L)$ obtained from the overall spectrum.  This disagreement is introduced artificially by the identification method employed in this section.  In this identification rule, a pair of energy levels accumulated strongly is decomposed into different components.  For this reason, the singularity characterizing the statistical properties of the whole spectrum is suppressed, and the contribution to this singularity from the accumulated pairs is limited only to the contribution from the pair of levels which are identified with the same component.  
%
%
%
%
%
%
%
%
%-*-*-*-*-*-*-*-*-*-*-*-*-*-*-*-*-
%  Discussion and  Conclusion
%-*-*-*-*-*-*-*-*-*--*-*-*-*-*-*-*
%
\section{Discussion and conclusion}
\label{sect5}
In this paper, starting from the formula (\ref{1.2}), we investigated the energy level statistics of bounded quantum system with infinitely many components and studied its deviations from Poisson statistics.  In the semiclassical limit, reflecting infinitely fine 
phase space structures of the classical dynamical system, individual energy eigenfunctions are expected to be localized in the phase space and to contribute independently to the level statistics.  With this in mind, we considered the situation in which the system consists of infinitely many components, each of which gives an infinitesimal contribution.  Then by applying the arguments of Pandey\cite{[2]} and Berry and Robnik,\cite{[0]} the level number variance, the spectral rigidity and the two-level cluster function were obtained within the limit of infinitely many spectral components.  These limiting observables are described by a single function, $\bar{c}(r)$, of the energy interval $r$, which is characterized by $\bar{c}(r)=0$ for the Poisson statistics and by $\bar{c}(r)\not=0$ when there are deviations from Poisson statistics.  Thus, even when the energy levels of individual components are statistically independent, non-Poissonian level statistics are possible.

Note that a singular two-level cluster function can be taken into account by a non-smooth cumulative function $c_i$.  Such a singularity is expected when there is strong clustering (accumulation) of the energy levels of individual components.  One possible example containing such strong clustering is the rectangular billiard system.  As shown in the section \ref{sect4}, this system exhibits deviations from Poisson statistics when the aspect ratio of the rectangle is a rational number.  To obtain a suitable method of dividing each energy level into the components, it is necessary to carry out further studies of this system.  Other examples have been studied by Shnirelman,\cite{[26]}\ Chirikov and Shepelyansky,\cite{[30]}\ and Frahm and Shepelyansky \cite{[31]} for certain types of systems containing quasi-degeneracy of levels resulting from time-reversal symmetry.  As is well known, the existence of quasi-degeneracy leads to non-Poissonian level statistics.  

It is also interesting to extend the Seligman-Verbaarschot formula\cite{[1]} for the level statistics of nearly integrable quantum systems with two degrees of freedom.  The classical phase space of a nearly integrable quantum system consists of regular and chaotic regions.  For this kind of system, Seligman and Verbaarschot proposed the number variance of the entire energy spectrum consisting of a subspectrum ($i=0$) exhibiting Poisson statistics as well as subspectra ($i=1,2,3,\cdots,N$) respectively exhibiting RMT:
\begin{equation}
\Sigma^2_{\mbox{\tiny SV}}(L)=\Sigma^2_{\mbox{\tiny{Poisson}}}(\rho_0 L) 
+ \sum_{i=1}^N \Sigma_{i,\mbox{\tiny{RMT}}}(\rho_i L),\label{5.1}
\end{equation}
where $\Sigma^2_{\mbox{\tiny{Poisson}}}(L)= L$ and 
$\Sigma^2_{i,\mbox{\tiny{RMT}}}$ is the number variance of the $i$th chaotic component obtained from the RMT.  Because the Liouville measure of a chaotic region is larger than zero, i.e., $\rho_i>0(i\geq 1)$,  Assumption (i) is invalid for this system.  However, the regular regions consist of infinitely many subsets, and the present approach is partially applicable to the spectral components corresponding to regular regions.  Along the line of the approach employed in section \ref{sect2}, the original proposal for the level number variance and the spectral rigidity by Seligman and Verbaarschot should be replaced by
\begin{equation}
\Sigma^2_{\bar{c}}(L)=\left( \rho_0 L -\int_0^{\rho_0 L} \bar{c}(r)dr \right)
+ \sum_{i=1}^N \Sigma^2_{i,\mbox{\tiny{RMT}}}(\rho_i L).
\label{5.2}
\end{equation}
When $\bar{c}(r)=0$ for all $r$, the new formula (\ref{5.2}) is reduced to that of the original proposal by Seligman and Verbaarschot (Formula {\ref{5.1}}).  In the case that $\bar{c}(r)\not=0$, on the other hand, the formula (\ref{5.2}) allows deviations from the Seligman-Verbaarschot formula.  Therefore, this result might represent an extension of the Seligman-Verbaarschot formalism proposed for nearly integrable quantum systems.
\section*{Acknowledgements}
The authors would like to thank Professor N.~Minami, Dr. Y.~Sakamoto, Professor M.~Robnik, and Professor A. ~Shudo for penetrating comments and discussion which helped the authors to deepen and clarify their own arguments.  The authors 
also thank the Yukawa Institute for Theoretical Physics at Kyoto University.  Discussions during the YITP workshop YITP-W-04-14 on "Quantum chaos: Present 
status of theory and experiment" were useful in completing this work.  This work is supported by a Grant-in-Aid for Scientific Research from the Ministry of Education, Culture, Sports, Science and Technology, Japan, for Young Scientists (B) (No. 15740244) to H. M.
\end{document}